\documentclass[11pt]{article}

\usepackage[final]{latex/acl}

\usepackage{times}
\usepackage{latexsym}

\usepackage[T1]{fontenc}

\usepackage[utf8]{inputenc}

\usepackage{microtype}

\usepackage{inconsolata}

\usepackage{graphicx}
\graphicspath{{../figures/}{figures/}}

\usepackage{amsmath}
\usepackage{amssymb}

\usepackage{multirow}   
\usepackage{makecell}   
\usepackage{array}      
\usepackage{booktabs}

\usepackage{tikz}
\usepackage{tikz-3dplot}
\usetikzlibrary{positioning, arrows.meta, shapes, shadows}
\usepackage{pgfplots}
\usepackage{enumitem}

\title{DSA-Tokenizer: Disentangled Semantic-Acoustic Tokenization \\via Flow Matching-based Hierarchical Fusion}

\author{
    \textbf{Hanlin ZHANG}\textsuperscript{1,*, $\ddagger$}, 
    \textbf{Daxin Tan}\textsuperscript{2,*}, 
    \textbf{Dehua Tao}\textsuperscript{2}, 
    \textbf{Xiao Chen}\textsuperscript{2, $\dagger$}, 
    \textbf{Haochen Tan}\textsuperscript{2}, \\ 
    \textbf{Yunhe Li}\textsuperscript{1}, 
    \textbf{Yuchen Cao}\textsuperscript{1}, 
    \textbf{Linqi Song}\textsuperscript{1,$\dagger$} \\
    \textsuperscript{1}Department of Computer Science, City University of Hong Kong \\
    \textsuperscript{2}AI Lab, Leibniz Research Center, Huawei \\
    \texttt{\{hanlzhang8-c@my., linqi.song@\}cityu.edu.hk}, \\
    \texttt{\{chen.xiao2, tan.daxin1\}@huawei.com}
}

\begin{document}

\maketitle

\begingroup
\renewcommand\thefootnote{}
\footnotetext{\textsuperscript{*} Equal contribution.}
\footnotetext{\textsuperscript{$\dagger$} Corresponding authors.}
\footnote{\textsuperscript{$\ddagger$}Work done during an internship at Huawei.}
\endgroup

\begin{abstract}
Speech tokenizers are a key building block of fully discrete Speech LLMs. 
Existing tokenizers either prioritize semantic encoding, fuse semantic content with acoustic style inseparably, or achieve incomplete semantic-acoustic disentanglement.
To achieve better disentanglement, we propose \textbf{DSA-Tokenizer}, 
which explicitly disentangles speech into discrete semantic and acoustic tokens via distinct optimization constraints. 
Specifically, semantic tokens are supervised by ASR to capture linguistic content, 
while acoustic tokens focus on mel-spectrograms restoration to encode style. 
We further introduce a hierarchical Flow Matching decoder and a joint reconstruction-context inpainting training strategy, 
allowing the model to support both high-fidelity reconstruction and cross-utterance voice clone. 
To speed up inference, we distill the DiT decoder to reduce sampling steps of inference to 4 and improve synthesis quality with GAN fine-tuning. 
Experiments demonstrate that DSA-Tokenizer provides strong semantic-acoustic disentanglement, 
reliable controllable voice cloning, 
and efficient high-fidelity generation with low WER/CER. 
Moreover, our results suggest that disentangled tokenization provides a more effective interface for downstream large-model speech generation.
Audio samples are avaialble at \url{https://anonymous.4open.science/w/DSA_Tokenizer_demo/}.

\end{abstract}

\section{Introduction}

The rapid advancement of large language models (LLMs) has catalyzed a paradigm shift in
speech processing, spawning Speech LLMs \citep{kimiaudio, MinMo, qwen2audio, qwen3omni}
that unify speech and language processing within a single framework. Among existing
architectures, \textbf{fully discrete Speech LLMs} \citep{zhang2025mimo, stepaudio,
glm4voice,nguyen2024spiritlminterleavedspoken} tokenize both input and output speech,
enabling end-to-end processing in a unified discrete space with seamless LLM
integration—yet their performance hinges heavily on the design of the speech tokenizer
\citep{survey}.

\begin{figure}[t]
  \includegraphics[width=\columnwidth]{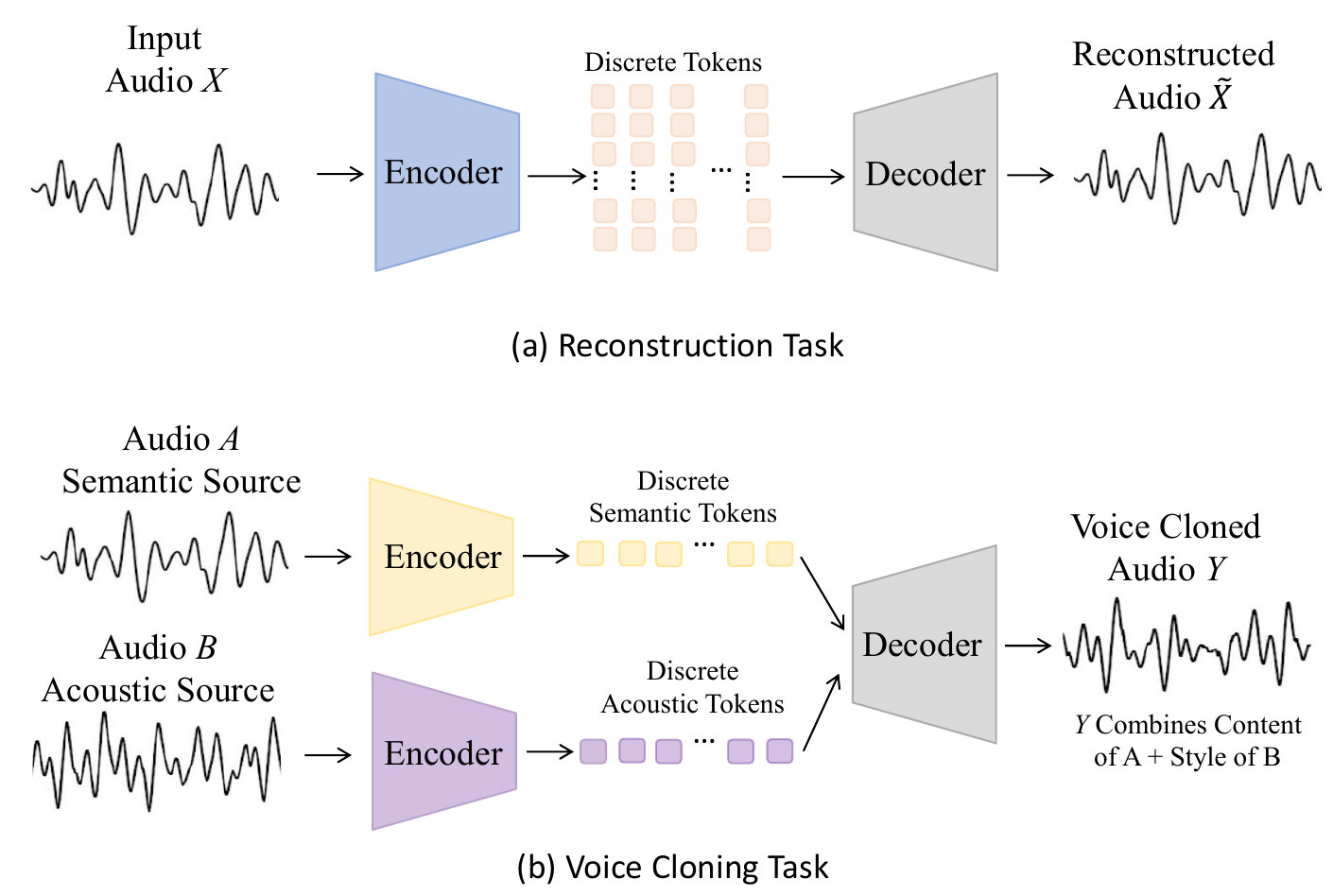}
  \caption{Illustration of (a) speech reconstruction and (b) cross-utterance voice cloning based on tokenizer}
  \label{fig:illustra}
\end{figure}

Existing speech tokenizers generally fall into three categories. \textbf{Semantic
tokenizers}~\citep{lakhotia2021generativespokenlanguagemodeling,HuBERT,cosyvoice2}
prioritize linguistic information via self-supervised learning or ASR supervision. While
this facilitates integration with LLMs, they often discard essential acoustic cues like
timbre. Conversely, \textbf{semantic-acoustic mixed tokenizers}~\citep{encodec,
wavtokenizer} target high-fidelity reconstruction but produce entangled representations,
preventing independent attribute control. Finally, \textbf{shallowly disentangled
tokenizers}~\citep{défossez2024moshispeechtextfoundationmodel, speechtokenizer} attempt
to decouple semantic and acoustic information atop mixed architectures. However, they
often suffer from incomplete disentanglement, failing to achieve a clean separation of
attributes.

To rigorously evaluate the disentanglement capability of these tokenizers, rather than
relying solely on reconstruction quality or ASR performance, we argue that
\textbf{cross-utterance voice cloning} (Figure \ref{fig:illustra}) is a more direct evaluation task. In this
setting, the semantic content is taken from one utterance, while the speaker and
prosodic style are taken from another. This requires the tokenizer to preserve
content and style independently, making semantic-acoustic leakage immediately observable.
Our experiments demonstrate that existing tokenizers exhibit substantial
limitations in this task. Some models bias towards linguistic information, causing
severe timbre mismatch, whereas others prioritize acoustic cues at the expense of
semantic fidelity. 

To address this gap, we propose \textbf{DSA-Tokenizer}. To ensure strict
disentanglement, we utilize a \textbf{constrained dual-stream tokenizer} where semantic
and acoustic tokens are supervised by ASR and a mel-spectrograms reconstruction objective,
respectively. These streams are processed by a \textbf{flow-matching-based hierarchical fusion
decoder}, which allows for high-fidelity reconstruction and cross-utterance
voice cloning free from rigid length constraints. Finally, by adopting a \textbf{joint
reconstruction-context inpainting training strategy} that combines self-reconstruction with
contextual inpainting, we enforce the robust separation of attributes necessary for
controllable speech generation. 

To further improve practicality, we distill the DiT-based decoder, enabling generation in just four sampling steps,
inspired by recent flow matching distillation approach ZipVoice\citep{zipvoice}.  
We then apply an additional GAN fine-tuning stage on the full mel-spectrogram 
generated by the distilled decoder. This design is motivated by 
prior observations that Flow Matching in mel space, when trained with 
a mean squared error (MSE) objective, tends to emphasize dominant coarse structures 
while under-modeling perceptually salient fine-grained details, especially in 
low-energy regions \citep{flow2gan}. This GAN fine-tuning stage 
substantially improves audio quality while preserving semantic-acoustic controllability.

Experiments show that DSA-Tokenizer achieves strong reconstruction, voice cloning, 
and efficient high-quality 4-step decoding through GAN refinement. 
Further results on downstream Speech LLM tasks
suggest that strict semantic-acoustic disentanglement provides a more
effective token interface for controllable speech generation.
Our core contributions are summarized as follows:

\begin{itemize}[leftmargin=*, nolistsep]
\item We propose \textbf{DSA-Tokenizer}, a dual-stream speech tokenizer that
explicitly disentangles semantic content and acoustic style while remaining
compatible with fully discrete Speech LLMs.

\item We introduce a DiT refinement strategy that distills the Flow-Matching decoder to perform inference in only 4 steps and improves mel‑spectrogram generation with GAN fine‑tuning, achieving a better audio generation quality.

\item Experiments on downstream large-model speech generation show that strict semantic-acoustic
disentanglement improves controllability and supports more effective speech generation.
\end{itemize}

\section{Related Work}
\subsection{Speech Large Language Models}
Recent Speech LLMs mainly follow two paradigms: thinker-talker systems that map
continuous speech to discrete outputs, and fully discrete systems that tokenize both
input and output speech \citep{MinMo, qwen3omni, glm4voice, zhang2025mimo,
nguyen2024spiritlminterleavedspoken}. Our work belongs to the latter line, where the
capability of downstream speech generation depends heavily on the quality of the
speech tokenizer.

\subsection{Disentangled Speech Tokenizers}

Existing speech tokenizers mainly trade off among semantic preservation,
reconstruction fidelity, and semantic-acoustic disentanglement. Semantic tokenizers
such as HuBERT\citep{HuBERT} approaches favor linguistic content but discard speaker and
prosodic style. Neural codecs such as EnCodec\citep{encodec} and 
WavTokenizer\citep{wavtokenizer} preserve rich acoustic
detail but entangle content and style in shared token streams. More recent
disentanglement-oriented methods, including SpeechTokenizer\citep{speechtokenizer}, 
DualCodec\citep{dualcodec}, XY-Tokenizer\citep{xytokenizer},
and SAC\citep{sac}, explicitly separate semantic and acoustic information through layered or
dual-branch designs. However, these approaches still suffer from incomplete
disentanglement, which limits their use in controllable speech generation.

\begin{figure*}[ht!]
    \centering
    \includegraphics[width=\linewidth]{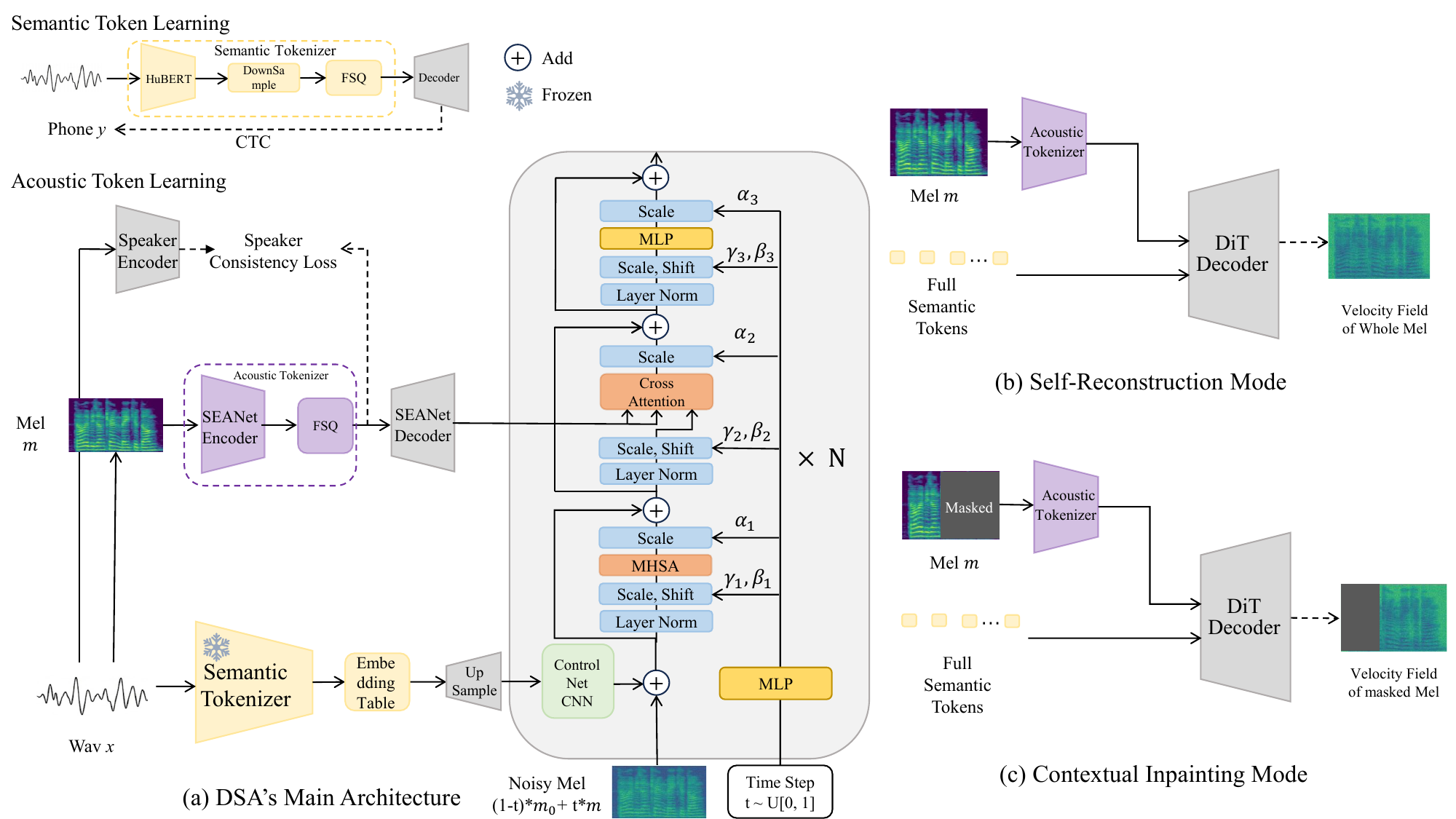}
    \caption{Overview of the proposed framework and training strategy. 
    (a) DSA-Tokenizer framework: Input audio $X$ is encoded into discrete semantic and acoustic tokens, which are fed into the DiT decoder for audio generation. 
    (b) Self-Reconstruction Mode: The model learns to predict the velocity field of the full Mel-spectrogram based on the complete acoustic and semantic tokens.
    (c) Contextual Inpainting Mode: The model learns to predict the velocity field of the masked Mel-spectrogram region based on the acoustic tokens of the unmasked region and the full semantic tokens.}
    \label{fig:architecture}
\end{figure*}

\section{Method}
We propose \textbf{Disentangled Semantic-Acoustic Tokenizer} (DSA-Tokenizer), 
a disentangled speech tokenization framework tailored for fully discrete Speech LLMs, as shown in
Figure\ref{fig:architecture}(a). Its goal is to achieve semantic-acoustic disentanglement 
for high-fidelity speech reconstruction and cross-utterance voice cloning. 
To this end, the framework employs two parallel discrete token streams with orthogonal goals: 
semantic tokens ($z_s$) encoding linguistic content, and acoustic tokens ($z_a$) capturing style
attributes. These two token streams are then fused via distinct condition injection
methods in a DiT-based Flow-Matching decoder. To further enable efficient 
high-quality generation, we distill the decoder to 4-step inference and refine it with adversarial 
fine-tuning.

\subsection{Dual Token Streams with Task-Specific Constraints}

\subsubsection{Semantic Token Learning}
Following recent discrete speech tokenization work \citep{chen2025emova,
tao2024toneunit, dstk2025}, we learn semantic tokens under ASR supervision so that
they retain linguistic content while suppressing acoustic style information.
Given a speech waveform $x$, we use a pre-trained HuBERT
model as the semantic encoder, followed by a Finite Scale Quantization (FSQ)
layer \citep{fsq} to discretize the continuous representations into semantic tokens
at 25 Hz:
\[
z_s = \text{FSQ}\left(\text{HuBERT}(x)\right) \in \mathcal{Z}^{T_s \times D_s}
\]
where $\mathcal{Z}$ denotes the discrete token space, $T_s$ is the sequence length, and
$D_s$ is the FSQ codebook size. We enforce strict linguistic constraints by training the
HuBERT-FSQ pipeline with Connectionist Temporal Classification (CTC) loss
\citep{graves2006connectionist}. After training, the auxiliary CTC decoder 
is discarded and the HuBERT-FSQ module is frozen as the semantic tokenizer 
for subsequent acoustic token learning. The semantic tokenizer uses FSQ with
6 channels and 4 levels per channel ($4^6$).

\subsubsection{Acoustic Token Learning}

Acoustic tokens are designed to capture speaker and prosodic attributes that are 
complementary to the linguistic content in semantic tokens.
Specifically, mel-spectrograms $m$ are extracted from raw speech $x$, downsampled via a
SEANet-style \citep{SEANet} encoder, and discretized through a FSQ quantizer to yield
the acoustic tokens:
\[
z_a = \text{FSQ}\left(\text{SEANetEncoder}(m)\right) \in \mathcal{Z}^{T_a \times D_a}
\]
where $T_a$ denotes the sequence length jointly determined by the CNN structure and
mel-spectrogram length, and $D_a$ is the FSQ codebook size. This design does not require $T_a$ and $T_s$ 
to have the same length, facilitating cross-utterance voice cloning.

Unlike the pre-trained semantic token stream, the acoustic tokenizer is trained
\textbf{end-to-end} with the DiT decoder (Sec.~\ref{sec:flow_matching}). The Flow
Matching loss is backpropagated through the discrete bottleneck via the
straight-through estimator to optimize the SEANet encoder. This drives $z_a$ to
capture residual acoustic detail beyond the linguistic content encoded in $z_s$.
The acoustic tokenizer uses FSQ with 8 channels and 4 levels per channel
($4^8$).

\subsubsection{Joint Reconstruction-Contextual Inpainting Training Strategy}

Training with reconstruction alone often encourages information leakage between
semantic and acoustic representations and makes voice cloning harder to
support. To address this issue, we adopt a joint \textbf{reconstruction-contextual inpainting} 
training strategy, where each batch is randomly assigned to one of two modes with equal probability. 

\noindent \textbf{Self-Reconstruction Mode} (Figure \ref{fig:architecture}(b)): The full
sequences of $z_s$ (semantic tokens) and $z_a$ (acoustic tokens) are provided as
conditions. The model learns to predict the Flow Matching velocity field of the entire
mel-spectrogram for high-fidelity reconstruction.

\noindent \textbf{Contextual Inpainting Mode} (Figure
\ref{fig:architecture}(c)): we randomly sample a split point $\tau$ on the
time axis and mask the mel-spectrogram after $\tau$. The decoder is then trained to
predict the masked region conditioned on the full semantic token sequence $z_s$ and
only the acoustic prefix $z_a^{<\tau}$. This design encourages the model to infer
global acoustic style from partial acoustic context while following semantic guidance,
thereby reducing information leakage between the two streams.

\subsection{Hierarchical Flow Matching Decoder for Semantic-Acoustic Token Fusion \label{sec:flow_matching}}
Fusing semantic tokens ($z_s$) and acoustic tokens ($z_a$) requires modeling two
fundamentally different types of conditions: linguistic content demands strict temporal
alignment, whereas acoustic style should be injected more flexibly across time. We
therefore adopt a DiT-based \citep{dit, ho2020denoising} Flow Matching
\citep{flowmatching} decoder with a \textbf{hybrid injection strategy}. Semantic
tokens are treated as the structural backbone of speech and injected as dense temporal
conditions to enforce content alignment. Acoustic tokens, in contrast, are injected via
\textbf{cross-attention}, allowing the decoder to retrieve speaker and prosodic cues. This design enables both high-fidelity
reconstruction and cross-utterance voice cloning.

\subsubsection{Semantic ControlNet-Style Injection}

We first project semantic tokens $z_s$ into continuous embeddings $e_s$ via a
learnable codebook, and then upsample them with linear interpolation to match the
mel-spectrogram length $T_{\text{mel}}$, yielding aligned semantic features
$\tilde{e}_s$.
Following standard Flow Matching paradigms, at timestep $t \sim \mathrm{U}[0,1]$, 
the noisy mel-spectrogram is defined as
\[
m_t = (1-t)\cdot m_0 + t\cdot m,
\]
where $m_0 \sim \mathcal{N}(0,I)$ and $m$ is the target mel-spectrogram. To enforce
strict linguistic alignment, we inject semantic information as a dense structural
condition. Specifically, a lightweight ControlNet-Style\citep{controlnet} CNN adapter 
processes $\tilde{e}_s$, and the result is directly added to $m_t$:
\[
F_{\text{sem}} = \operatorname{CNN}(\tilde{e}_s) + m_t.
\]
By treating semantic information as the structural backbone embedded into the noisy input, we ensure
the model captures long-range linguistic context prior to integrating acoustic details.


\subsubsection{Acoustic Injection via Cross-Attention}
Unlike semantic content, acoustic style does not require strict frame-level
alignment. We therefore inject acoustic information through cross-attention to
provide flexible speaker and prosodic conditioning. Specifically, acoustic
tokens $z_a$ are projected to continuous embeddings via a learnable codebook and
upsampled to the mel-spectrogram length $T_{\text{mel}}$ using a SEANet-style decoder,
yielding the aligned acoustic features $\tilde{e}_a$.

Within the DiT blocks, the self-attention-enhanced semantic feature $F_{\text{sem}}$
serves as queries, while keys and values are derived from $\tilde{e}_a$ via dedicated
linear projections $W^K$ and $W^V$. The cross-attention fused feature is computed as:
\[
F_{\text{sem-aco}} = \text{CrossAttn}\left(F_{\text{sem}}, \tilde{e}_a W^K, \tilde{e}_a
W^V\right)
\]
This design allows the decoder to retrieve speaker and prosodic cues flexibly
without imposing rigid length constraints between semantic and acoustic tokens, 
thus facilitating cross-utterance voice cloning.

\subsubsection{Training Objectives}
We train the base decoder with a combination of Flow Matching loss for conditional
mel-spectrogram generation and speaker consistency loss for acoustic style
preservation.

\paragraph{Flow Matching Loss} 

Following the Conditional Flow Matching (CFM) \citep{cfm}, given the noisy mel-spectrogram
\[
m_t = (1-t)m_0 + tm,
\]
where $m_0 \sim \mathcal{N}(0, I)$ and $t \sim \mathrm{U}[0,1]$, we train the decoder
to predict the corresponding velocity field conditioned on the semantic and acoustic
features:
\[
\mathcal{L}_{\text{fm}} = \mathbb{E}_{t,m_0,m}
\left[
\left\|
v_t - v_\theta(m_t, t, \tilde{e}_s, \tilde{e}_a)
\right\|^2
\right].
\]

\paragraph{Speaker Consistency Loss} 

To encourage acoustic tokens to preserve speaker identity during the first-stage
training, we extract a reference speaker embedding $s_{\text{ref}}$ from the raw
waveform using a WavLM-based speaker verification model \citep{wavlm}
\footnote{\url{https://github.com/microsoft/UniSpeech/tree/main/downstreams/speaker_verification}\label{fn:speaker}}.
We then align it with the pooled acoustic-token representation:
\[
\mathcal{L}_{\text{spk}} = 1 - \cos\left(s_{\text{ref}},
\operatorname{AttnPool}({e}_a)\right)
\]
where $\operatorname{AttnPool}(\cdot)$ aggregates the acoustic token embeddings
$\mathbf{e}_a$ into a global vector. 

The total training loss is a weighted sum:
$\mathcal{L}_{\text{total}} = \mathcal{L}_{\text{fm}} + \lambda_{\text{spk}}
\mathcal{L}_{\text{spk}}$, where $\lambda_{\text{spk}}=1.0$. Additional objectives used in few-step
distillation and adversarial fine-tuning are introduced in the next subsection.

\subsubsection{Few-step distillation and Adversarial Fine-Tuning}
To improve decoding efficiency and make the end-to-end GAN fine-tuning feasible, 
we freeze the tokenizer and distill the pretrained Flow-Matching decoder into a 
4-step student following ZipVoice.
During distillation, each batch retains both self-reconstruction and contextual
inpainting modes, and the student DiT decoder is trained on forward passes from
both, allowing the student to preserve reconstruction fidelity and cross-utterance
voice cloning capability under short-step decoding.

We then adversarially fine-tune the distilled 4-step model in mel space, while
keeping the tokenizer frozen. For each batch, self-reconstruction and contextual
inpainting samples are both forwarded for 4 steps to generate full
mel-spectrograms. Their effective mel lengths vary across samples: self-reconstruction
examples inherit the original utterance duration, while the supervised region in
contextual inpainting depends on the randomly sampled mask boundary. We therefore
define the adversarial training length as the shortest effective mel length $t$
within the batch, and randomly crop one segment of length $t$ from each
fake/real pair before feeding it to a multi-scale mel-domain PatchGAN\citep{patchgan}
discriminator. This design enables stable joint adversarial training on mixed
batches without introducing additional length constraints.

To preserve speaker identity during this stage, we introduce a proxy speaker
encoder trained to match the embedding space of a frozen WavLM-based speaker
verification model\textsuperscript{\ref{fn:speaker}}. The proxy encoder operates
directly on generated mel-spectrograms and provides a differentiable mel-level
speaker consistency loss, which is applied only to self-reconstruction samples:
\[
\mathcal{L}_{\text{spk}}^{\text{mel}} =
1 - \cos\left(h_{\phi}(\hat{m}), s_{\text{ref}}\right),
\]
where $\hat{m}$ is the mel-spectrogram generated by the distilled decoder,
$h_{\phi}(\cdot)$ is the proxy speaker encoder, and $s_{\text{ref}}$ is the same
reference speaker embedding used in the first-stage speaker consistency loss. The
generator is optimized with the original first-stage objective together with
least-squares adversarial\citep{lsgan} and feature-matching losses\citep{featurematching}:
\[
\mathcal{L}_{G} =
\mathcal{L}_{\text{total}}
+ \lambda_{\text{adv}} \mathcal{L}_{\text{adv}}^{\text{mel}}
+ \lambda_{\text{feat}} \mathcal{L}_{\text{feat}}^{\text{mel}}
+ \lambda_{\text{spk}}^{\text{mel}} \mathcal{L}_{\text{spk}}^{\text{mel}}.
\]
More training details are provided in Appendix~\ref{gan-finetuning}.

\section{Experiment Setup}

\subsection{Evaluation Tasks}

\subsubsection{Reconstruction and Cross-Utterance Voice Cloning}
We evaluate DSA-Tokenizer on two waveform-generation settings:
\textbf{Reconstruction}, where a single utterance provides both semantic and
acoustic sources, and \textbf{Cross-Utterance Voice Cloning}, where semantic
and acoustic tokens are taken from different utterances. We report UTMOS
\citep{utmos} for speech naturalness, WER for English and CER for Chinese for
content consistency, and SIM for speaker/style preservation.

\subsubsection{Disentanglement Probing}
To assess semantic-acoustic disentanglement, we probe whether different token
streams preserve only task-relevant information. ASR is used to evaluate
semantic information, and speaker classification (SC) is used to evaluate
speaker-related acoustic information. For single-layer baselines, ASR and SC are
applied to the same token sequence; for multi-layer baselines, both are applied
to Layer 0 and Layers 1--7; for DSA-Tokenizer and SAC, both are applied separately to semantic and acoustic tokens. For Facodec, ASR is applied to content code and SC is applied to other parts of Facodec.

\subsubsection{Downstream Speech LLM Experiments}
We further evaluate whether DSA-Tokenizer provides a better token interface for
speech LLMs through two downstream tasks: \textbf{LLM-based voice cloning} and
\textbf{LLM-based TTS}. For voice cloning, the autoregressive model is trained
on triplets $(\mathcal{S}, \mathcal{A}, \mathcal{C})$, where $\mathcal{S}$
provides semantic content, $\mathcal{A}$ provides acoustic style, and
$\mathcal{C}$ is the target speech. For zero-shot TTS, we adopt a pure LLaSA-style autoregressive architecture \citep{llasa}, which generates speech tokens
conditioned on text and prompt speech.

\subsection{Datasets}
\label{app:datasets}
\noindent\textbf{DSA-Tokenizer Training:}
We train the semantic tokenizer on about 4,000 hours of Chinese-English
speech-text aligned data, and train the acoustic tokenizer and decoder on a
cleaned 100k-hour Chinese-English subset of Emilia \citep{emilia}. Additional
training, cleaning, and architecture details are provided in
Appendices~\ref{cleaning dataset}, \ref{training detail}, and \ref{model_arch}.

\noindent\textbf{Waveform Evaluation:}
We use the multilingual SeedTTS dataset \citep{seedtts} for English and Chinese
reconstruction and voice cloning evaluation.

\noindent\textbf{Disentanglement Probing:}
We use LibriSpeech \citep{librispeech} for ASR-based semantic evaluation and
VoxCeleb1 \cite{nagrani2020voxceleb} for speaker classification.

\noindent\textbf{LLM-based Voice Cloning:}
We construct over 350,000 training triplets from LibriTTS train-960 and
evaluate on over 4,000 randomly matched test pairs from LibriTTS test-clean.
Target speech is synthesized by F5TTS \citep{f5tts} and Cosyvoice2 \citep{cosyvoice2}.

\noindent\textbf{LLM-based TTS:}
We follow the same data distribution as VoxBox \citep{sparktts}.

\subsection{Baselines}
We compare against representative single-layer, multi-layer, and dual-stream
tokenizers, including WavTokenizer, Mimi, EnCodec, SpeechTokenizer, DualCodec,
SAC, Cosy2 $S^3$, and Facodec. 
For both LLM-based voice cloning and TTS, we use
Qwen3-0.6B \citep{qwen3} as the autoregressive backbone and expand its
vocabulary with the corresponding speech tokens. Additional dataset,
task-formatting, and baseline details are provided in
Appendices~\ref{app:exp_setup_details} and \ref{app:baseline_details}.

\section{Results and Discussion}
\subsection{Can DSA-Tokenizer Unify Reconstruction and Voice Cloning?}
\begin{table*}[ht!]  
\centering
\small  
\renewcommand{\tabcolsep}{1.0mm} 
\begin{tabular}{lcccccccc}
\toprule
\multirow{2}{*}{\textbf{Model}} & 
\multirow{2}{*}{\makecell[cc]{BitRate \\ (kbps)}} & 
\multicolumn{3}{c}{\textbf{English}} & 
\multicolumn{3}{c}{\textbf{Chinese}} \\
\cmidrule(lr){3-5} \cmidrule(lr){6-8}
& & UTMOS $\boldsymbol{\uparrow}$ & WER (\%) $\boldsymbol{\downarrow}$ & SIM $\boldsymbol{\uparrow}$ & UTMOS $\boldsymbol{\uparrow}$ & CER (\%) $\boldsymbol{\downarrow}$ & SIM $\boldsymbol{\uparrow}$ \\
\midrule
Ground Truth      & -  & - & 2.14 & - & - & 1.33 & - \\
\midrule
\multicolumn{8}{l}{\textit{\textbf{Reconstruction Task}}} \\
WavTokenizer (75 Hz)       & 0.90  & 3.92 & 3.23 & \textbf{0.84} & 2.82 & 5.08 & 0.62 \\
Mimi (8-layer)            & 1.10  & 3.30 & 3.46 & 0.74 & 2.37 & 2.93 & 0.73 \\
Encodec (2-layer)         & 1.50  & 1.56 & 5.40 & 0.61 & 1.35 & 5.19 & 0.62 \\
SpeechTokenizer (2-layer) & 1.00  & 2.12 & 7.57 & 0.36 & 1.75 & 24.39 & 0.33 \\
DualCodec (12.5 Hz, 6-layer) & 0.925  & 3.78 & 2.58 & 0.76 & 2.95 & 1.62 & 0.80 \\
SAC ($f_s$=12.5 Hz, $f_a$=50 Hz) & 0.875 & 3.88 & 2.03 & 0.83 & 2.99 & \textbf{1.53} & \textbf{0.87}\\
Facodec & - & 3.54 &2.26 &0.84 & 2.82 & 2.01 & 0.83 \\
\textbf{DSA-Tokenizer ($f_s$=25 Hz, $f_a$=25 Hz)} & \textbf{0.70}  & \textbf{4.06} & \textbf{2.00} & 0.77 & \textbf{3.30} & 1.81 & 0.82 \\
\midrule
\multicolumn{8}{l}{\textit{\textbf{Cross-Utterance voice cloning Task}}} \\
Mimi (8-layer)            & 1.10  & 2.19 & 107.51 & 0.53 & 1.54 & 100.01 & 0.53 \\
Encodec (2-layer)         & 1.50  & 1.25 & 98.31  & 0.09 & 1.24 & 68.97  & 0.13 \\
SpeechTokenizer (2-layer) & 1.00  & 1.40 & 12.98  & 0.13 & 1.34 & 123.33 & 0.15 \\
DualCodec (12.5 Hz, 6-layer) & 0.925  & 2.39 & 17.36  & 0.11 & 1.73 & 16.22   & 0.32 \\
SAC ($f_s$=12.5 Hz, $f_a$=50 Hz) & 0.875 & 1.34 & 90.22  & 0.13 & 1.27 & 71.87  & 0.30 \\
Facodec & - & 2.52 & 5.78 & 0.26 & 1.86 &25.1 &0.39 \\
\textbf{DSA-Tokenizer ($f_s$=25 Hz, $f_a$=25 Hz)} & \textbf{0.70}  & \textbf{4.16} & \textbf{2.47}  & \textbf{0.61} & \textbf{3.55} & \textbf{2.16}  & \textbf{0.71} \\
\bottomrule
\end{tabular}
\renewcommand{\tabcolsep}{6pt}  
\caption{Performance comparison of different tokenizers on speech reconstruction and cross-utterance voice cloning tasks. $f_s$ means semantic token rate and $f_a$ means acoustic token rate. Baseline models with similar bitrate are selected,annotated with their codebook
layers or token rates.}
\label{tab:recon_recomb_combined}
\end{table*}
\textit{DSA-Tokenizer is the only tokenizer that remains strong in both
high-fidelity reconstruction and cross-utterance voice cloning.} As shown in
Table~\ref{tab:recon_recomb_combined}, DSA-Tokenizer achieves the highest UTMOS
scores in both English and Chinese under both evaluation settings, while also
maintaining very low WER/CER at the lowest bitrate. 
In reconstruction task, although DSA-Tokenizer does not achieve the highest SIM, we argue it is a \textbf{trade-off} under a tight bitrate budget: instead of maximizing reconstruction SIM, 
our model learns more usable semantic-acoustic tokens that 
better support Speech LLMs(Section~\ref{sec:downstream_applications}).
For voice cloning task, DSA-Tokenizer achieves the best UTMOS, SIM, and WER/CER simultaneously. 
In contrast, reconstruction-oriented non-disentangled baselines such as
WavTokenizer can achieve strong reconstruction quality, but they do not support
robust voice cloning. Among disentangled baselines, SAC remains
competitive in reconstruction, yet suffers a severe degradation in voice cloning.
Although FaCodec uses continuous timbre features, it still performs consistently worse than DSA-Tokenizer across both languages.
We further compare DSA-Tokenizer with the CosyVoice series, which is not fully discrete, 
for voice cloning in Appendix~\ref{app:cosyvoice_compare}. 
DSA-Tokenizer outperforms all
CosyVoice systems in English and achieves the best Chinese CER and UTMOS while remaining
competitive in SIM.

\subsection{Can Discrete Tokens Capture Sufficient Information without Leakage?}\label{sec:disentanglement_probing}
\begin{figure}[t]
  \centering
    \includegraphics[width=\columnwidth]{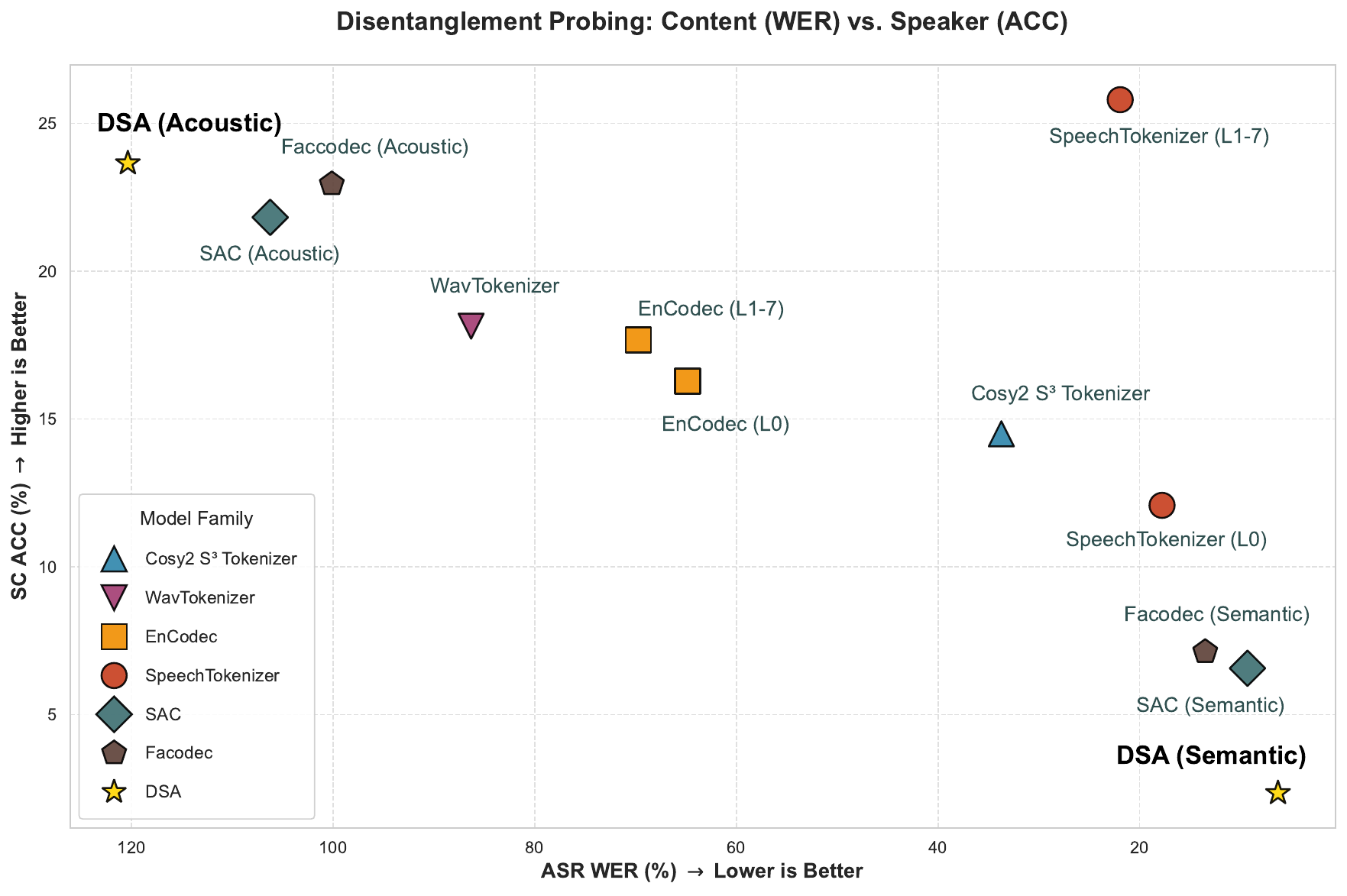}
  \caption{Disentanglement probing evaluation results. (L0) means the first layer, (L1-7) means the second to eighth layers. Tokens from multiple layers are concatenated for the experiments.}
  \label{fig:probing}
\end{figure}
\textit{DSA-Tokenizer preserves sufficient task-relevant 
information in both kinds of tokens while minimizing leakage.} As
illustrated in Figure~\ref{fig:probing}, different tokenizers exhibit distinct encoding
patterns based on their objectives. First, Single-codebook models are highly dependent 
on their training objectives: the ASR-supervised CosyVoice2 $S^3$ tokenizer is 
more semantic-oriented, whereas the reconstruction-trained WavTokenizer retains more 
speaker-related information at the cost of much worse WER. Second,
multi-layer tokenizers demonstrate varying disentanglement capabilities. EnCodec,
lacking explicit constraints, shows uniform performance across layers, suggesting
semantic and acoustic entangled. In contrast, SpeechTokenizer achieves shallow
disentanglement by distilling semantics into layer 0, which yields significantly lower
WER and SC ACC compared to layers 1–7. Third, among dual-stream baselines, both SAC and
Facodec achieve partial disentanglement, but their semantic tokens still retain
more speaker information and their acoustic tokens retain more semantic residue
than DSA-Tokenizer. In contrast, DSA-Tokenizer attains the lowest WER on
semantic tokens (6.28\%) together with the lowest SC ACC (2.35\%), while its
acoustic tokens achieve the highest SC ACC (23.65\%) and the highest WER
(120.36\%). These results suggest that DSA-Tokenizer provides the cleanest
semantic-acoustic disentanglement among the compared tokenizers. This probing pattern is also
consistent with its superior voice cloning performance in
Table~\ref{tab:llm_vc}. Taken together, the probing and voice cloning results suggest that
voice cloning is a \textbf{stringent test} of semantic-acoustic disentanglement, 
and that cleaner disentanglement in tokenizers leads to 
more effective voice cloning.

\subsection{Does Disentanglement Benefit Downstream Speech LLMs?}
\label{sec:downstream_applications}
\textit{Downstream speech LLM experiments suggest that cleaner
semantic-acoustic disentanglement provides a more effective token interface for
acoustic-related generation.} 
In LLM-based voice cloning
(Table~\ref{tab:llm_vc}), the non-disentangled WavTokenizer performs worst on
all metrics and even leads to unstable generation, showing that strong
reconstruction alone does not guarantee robust downstream modeling. Compared
with the partially disentangled SAC, DSA-Tokenizer achieves clearly better
UTMOS and SIM while also improving WER, indicating that cleaner factorization
makes cross-utterance content-style voice cloning easier for the LLM.

This trend remains visible in pure LLaSA-style autoregressive TTS
(Table~\ref{tab:llm_tts}). Non-disentangled tokenizers used by the LLaSA
baselines and Spark-TTS can produce reasonable speech, but they remain weaker
than disentangled tokenizers in the overall performance.
Among disentangled tokenizers, SAC attains the lowest WER, suggesting that
partial disentanglement already helps content modeling in this setting. DSA-Tokenizer,
however, achieves the best UTMOS and SIM in both English and Chinese while
keeping WER competitive.
Taken together, these results suggest that \textbf{cleaner semantic-acoustic
disentanglement provides a better token interface for Speech LLMs.}

\begin{table}[t]
\centering
\resizebox{\linewidth}{!}{
\begin{tabular}{lccc}
\toprule 
Tokenizer & UTMOS $\uparrow$ & WER (\%) $\downarrow$ & SIM $\uparrow$ \\
\midrule 
WavTokenizer & 3.75 & 89.44 & 0.28 \\
SAC & 3.35 & 22.89 & 0.38 \\
\textbf{DSA-Tokenizer} & \textbf{4.10} & \textbf{16.55} & \textbf{0.48} \\
\bottomrule 
\end{tabular}
}
\caption{LLM-based voice cloning performance.}
\label{tab:llm_vc}
\end{table}

\begin{table}[t]
\centering
\small
\begin{tabular}{lccc}
\toprule
\textbf{Model} & WER (\%) $\downarrow$ & SIM $\uparrow$ & UTMOS $\uparrow$ \\
\midrule
\multicolumn{4}{l}{\textit{\textbf{Seed-TTS test-en}}} \\
Llasa-1B-80k & 3.71 & 0.54 & 4.06 \\
Llasa-1B-160k & 3.60 & 0.56 & 4.05 \\
Llasa-1B-250k & 2.99 & 0.57 & 4.07 \\
Spark-TTS & 1.98 & 0.58 & 3.94 \\
SAC & \textbf{1.06} & 0.54 & 4.21 \\
\textbf{DSA} & 1.39 & \textbf{0.60} & \textbf{4.24} \\
\midrule
\multicolumn{4}{l}{\textit{\textbf{Seed-TTS test-zh}}} \\
Llasa-1B-80k & 2.69 & 0.65 & 3.27 \\
Llasa-1B-160k & 2.22 & 0.66 & 3.28 \\
Llasa-1B-250k & 1.89 & 0.67 & 3.28 \\
Spark-TTS & 1.20 & 0.67 & 3.27 \\
SAC & \textbf{0.90} & 0.65 & 3.34 \\
\textbf{DSA} & 1.37 & \textbf{0.68} & \textbf{3.51} \\
\bottomrule
\end{tabular}
\caption{Pure LLaSA-style autoregressive TTS performance on Seed-TTS.}
\label{tab:llm_tts}
\end{table}

\subsection{What Design Choices Are Essential?}

\begin{table}[t]
\centering
\resizebox{\linewidth}{!}{
\begin{tabular}{lcccccc}
\toprule 
& \multicolumn{3}{c}{\textbf{Reconstruction}} & \multicolumn{3}{c}{\textbf{Voice Cloning}} \\
\cmidrule(lr){2-4} \cmidrule(lr){5-7} 
\textbf{Model} & UTMOS & WER & SIM  & UTMOS & WER & SIM \\
\midrule 
DSA-Tokenizer        &  3.78 & 2.16 & 0.73 & 4.03 & 2.58 & 0.60 \\
w/o $\mathcal{L}_{\text{spk}}$ & 3.81 & 2.16 & 0.47 & 3.83 & 7.83 & 0.20 \\
w/o Contextual Inpainting     & 3.80 & 2.25 & 0.71 & 2.33 & 134.85 & 0.21 \\
\bottomrule 
\end{tabular}
}
\caption{Ablation study of speaker loss and contextual inpainting mode.}
\label{tab:ablation}
\end{table}

\textit{We ablate both the base tokenizer design and the short-step decoder
refinement.} We first examine the core design choices of the original
DSA-Tokenizer (before distillation and GAN fine-tuning) in
Table~\ref{tab:ablation}. Removing the speaker loss
($\mathcal{L}_{\text{spk}}$) has little effect on reconstruction UTMOS and WER, but causes a substantial drop in SIM. Its impact is much larger in voice
cloning, where UTMOS, WER, and SIM all degrade markedly, showing that
$\mathcal{L}_{\text{spk}}$ is important not only for preserving speaker
identity but also for stable cross-utterance style transfer. Removing
contextual inpainting has only a mild effect on reconstruction, but causes
voice cloning to collapse, showing that contextual inpainting is the key training mode
that enables robust voice cloning.

We further study the second-stage decoder refinement in
Table~\ref{tab:decoder_ablation}. Distilling the original decoder from 16 steps
to 4 steps preserves the overall reconstruction and voice cloning performance
remarkably well: UTMOS and WER remain highly comparable to the original model,
with only a moderate drop in SIM. Applying GAN fine-tuning on top of the distilled 
decoder further improves all metrics across both tasks, yielding the best UTMOS, 
WER, and SIM. In particular, the gain is
not limited to perceptual quality: content-related metrics are also improved,
indicating that adversarial refinement enhances synthesis fidelity without
weakening semantic-acoustic controllability.

\begin{table}[t]
\centering
\resizebox{\linewidth}{!}{
\begin{tabular}{lccccccc}
\toprule
\textbf{Model} & \textbf{Steps}
& \multicolumn{3}{c}{\textbf{Reconstruction}}
& \multicolumn{3}{c}{\textbf{Voice Cloning}} \\
\cmidrule(lr){3-5} \cmidrule(lr){6-8}
& & UTMOS & WER & SIM & UTMOS & WER & SIM \\
\midrule
Original Decoder         & 16 & 3.78 & 2.16 & 0.73 & 4.03 & 2.58 & 0.60  \\
4-step Distilled         & 4  & 3.87 & 2.18 & 0.70 & 4.04 & 2.49 & 0.56 \\
4-step Distilled + GAN   & 4  &\textbf{4.06} & \textbf{2.00} & \textbf{0.77}  & \textbf{4.16} & \textbf{2.47}  & \textbf{0.61} \\
\bottomrule
\end{tabular}
}
\caption{Ablation of short-step distillation and GAN refinement.}
\label{tab:decoder_ablation}
\end{table}

\section{Conclusion}

We present DSA-Tokenizer, a dual-stream speech tokenizer that explicitly
disentangles speech into discrete semantic and acoustic tokens. With
task-specific supervision, contextual inpainting training, 
DSA-Tokenizer supports both high-fidelity
reconstruction and cross-utterance voice cloning. We further distill the
decoder to 4-step inference and refine it with GAN fine-tuning for efficient
high-quality generation. Experiments show that DSA-Tokenizer provides the
cleanest semantic-acoustic disentanglement, the best overall balance between
reconstruction and voice cloning at low bitrate, and a more effective token
interface for Speech LLMs.

\section*{Limitations}

Despite its strong performance, DSA-Tokenizer still has several limitations.
First, although decoder distillation reduces inference to 4 steps, the model is
still not as lightweight as one-shot codec or GAN-based generators, which may
limit deployment in latency-sensitive settings. Second, our training and
evaluation focus on Chinese-English speech, cross-utterance voice cloning, and
LLM-based TTS; broader language coverage and more diverse downstream speech LLM
tasks remain unexplored. Third, our disentanglement analysis mainly probes
semantic information and speaker-related acoustic information through ASR and
speaker classification, and does not fully measure other factors such as
prosody or emotion. We leave these directions to future work.

\bibliography{latex/custom}

\appendix

\section{Model Architecture}
\label{model_arch}

\subsection{SEANetEncoder}
\label{SEANetEncoder}
For the SEANetEncoder, we employ a four-layer architecture with dimensions of [512,
1024, 1024, 1024]. The downsampling ratios are set to [2, 1, 1] and [2, 2, 1] for
acoustic token rates of 50Hz and 25Hz, respectively.

\subsection{SEANetDecoder}
\label{SEANetDecoder}
For the SEANetDecoder, we utilize a four-layer architecture with dimensions of [1024,
1024, 1024, 1024]. The upsampling ratios are set to [1, 1, 2] and [1, 2, 2] for acoustic
token rates of 50Hz and 25Hz, respectively.

\subsection{DiT blocks}

For the DSA-Tokenizer, we utilize a stack of $22$ DiT blocks with a hidden dimension of
$1024$. We incorporate AdaNorm~\citep{adanorm} for both multi-head self-attention and
multi-head cross-attention mechanisms. The FFN consists of two linear layers with an
inner dimension set to $4096$. And RoPE\citep{rope} is set as the position embedding for
both multi-head self-attention and multi-head cross-attention.

\subsubsection{ControlNet-Style CNN Adapter}

For the ControlNet-Style CNN Adapter, we employ a lightweight stack of three 1D
convolutional layers to extract contextual information while projecting the feature
dimension from 512 to 1024, as the embedding dim of semantic token is 512. Each layer
utilizes a kernel size of $3$ and a padding of $1$ to preserve the temporal sequence
length.

\subsection{Attention pooling layer}

Let $e_a \in \mathbb{R}^{T \times D}$ denote the input acoustic sequence, where $T$ is
the sequence length and $D=1024$ is the feature dimension. To aggregate the
variable-length sequence into a fixed-size embedding, we employ a Masked Attentive
Statistics Pooling (ASP) layer followed by a projection adapter.

First, the input sequence is transposed to $\mathbf{X} \in \mathbb{R}^{D \times T}$. A
channel-wise attention mechanism computes the importance scores $\mathbf{S} \in
\mathbb{R}^{D \times T}$ for each time step and feature dimension:
\[
\mathbf{S} = \mathbf{W}_2 \left( \tanh \left( \mathbf{W}_1 \mathbf{X} + \mathbf{b}_1
\right) \right) + \mathbf{b}_2,
\]
where $\mathbf{W}_1 \in \mathbb{R}^{D_{attn} \times D}$, $\mathbf{W}_2 \in \mathbb{R}^{D
\times D_{attn}}$ are convolution weights with kernel size 1, and $D_{attn}$ is the
bottleneck dimension, $\mathbf{b_1} \in \mathbb{R}^{D}$ and $\mathbf{b}_2 \in
\mathbb{R}^{D}$ are bias weights.

To handle variable sequence lengths, we apply a temporal mask $M \in \{0, -\infty\}^{T}$
based on the valid length of the sequence. The normalized attention weights
$\boldsymbol{\alpha} \in \mathbb{R}^{D \times T}$ are obtained via a masked softmax
operation along the temporal dimension:
\[
\alpha_{d,t} = \frac{\exp(S_{d,t} + M_t)}{\sum_{\tau=1}^{T} \exp(S_{d,\tau} + M_\tau)},
\]
where $d \in [1, D]$ and $t \in [1, T]$.

Using these weights, we calculate the weighted mean vector $\boldsymbol{\mu} \in
\mathbb{R}^{D}$ and the weighted standard deviation vector $\boldsymbol{\sigma} \in
\mathbb{R}^{D}$:
\begin{align*}
    \boldsymbol{\mu}_d &= \sum_{t=1}^{T} \alpha_{d,t} \cdot X_{d,t}, \\
    \boldsymbol{\sigma}_d &= \sqrt{\sum_{t=1}^{T} \alpha_{d,t} \cdot (X_{d,t} - \boldsymbol{\mu}_d)^2 + \epsilon}
\end{align*}
where $\epsilon$ is a small constant for numerical stability.

Finally, the statistics are concatenated and projected to the target dimension $D_{out}
= 256$:
\[
\mathbf{h} = \mathbf{W}_{proj} [\boldsymbol{\mu}; \boldsymbol{\sigma}] +
\mathbf{b}_{proj},
\]
where $[\cdot; \cdot]$ denotes concatenation resulting in a $2D$-dimensional vector
($2048$), and $\mathbf{W}_{proj} \in \mathbb{R}^{D_{out} \times 2D}$, $\mathbf{b}_{proj}
\in \mathbb{R}^{2D}$. The resulting $\mathbf{h}$ serves as the pooled acoustic token
embeddings.

\subsection{Proxy Speaker Encoder}

The encoder adopts a hybrid ECAPA-Transformer architecture\citep{ECAPA}. The input mel is first
normalized with InstanceNorm1d, followed by a Conv1d front-end and three SE-Res2 blocks
blocks with dilation rates 2, 3, and 4\citep{Res2Net}. Their outputs are concatenated and mixed by
1$\times$1 convolutions, and augmented with sinusoidal positional embeddings. 
We then apply 6 Transformer encoder blocks, and combine it with the ECAPA branch through a residual
connection. Finally, masked attentive statistics pooling produces a pooled feature. 

The Masked attentive statistics pooling is used to aggregate variable-length frame-level features into a fixed-dimensional utterance representation. Given the input sequence and its valid lengths, a temporal mask is first constructed to exclude padded frames from attention and pooling. The global mean and standard deviation over valid frames are computed and concatenated with the original features. The attention branch then applies two 1×1 convolution layers with a tanh activation to generate frame-wise attention weights, followed by masked softmax normalization over time. Finally, the module computes a weighted mean and weighted standard deviation and concatenates them to form the pooled feature.

\subsection{Codebook Size}

The acoustic tokenizer uses FSQ with 8 channels and 4 levels per channel,
corresponding to a codebook size of 65536. The semantic tokenizer uses FSQ 
with 6 channels and 4 levels per channel, corresponding to a codebook size of 4096.

\section{Dataset Cleaning}
\label{cleaning dataset}
We find that some samples of Emilia Dataset \citep{emilia} contains more than two
speakers. As we have contextual inpainting Mode, samples contains more
than two speakers will hurt the performance of our model, therefore, we utilize
speaker-diarization model
\footnote{\url{https://huggingface.co/pyannote/speaker-diarization}} \citep{ss} to
remove those dirty samples.

\section{Training Details}

\subsection{Training Details of DSA-Tokenizer}
\label{training detail}
All DSA-Tokenizer are trained witicdynamic batching with 30k frames of speech per batch.
We train the tokenizer for 400k steps using AdamW~\citep{adamw} with a learning rate of
$7.5 \times 10^{-5}$ and 32k warmup steps. The dataset used for the training of the
semantic tokenizer is sampled from LibriSpeech~\citep{librispeech},
GigaSpeech~\citep{chen2021gigaspeech}, Libri-Heavy~\citep{kang2024libriheavy},
CommonVoice~\citep{ardila2020common}, AISHELL-2~\citep{du2018aishell},
WenetSpeech~\citep{zhang2022wenetspeech}, MagicData-RAMC~\citep{yang2022open}, and
People's Speech~\citep{galvez2021people}. The DSA-Tokenizer has a parameter size of
approximately 430M. We trained the models on the Ascend platform for 15 days.

\subsection{Training Details of Short-Step Distillation and GAN Fine-Tuning}\label{gan-finetuning}
During both short-step distillation and GAN fine-tuning, the semantic and acoustic
tokenizers are frozen, and only the decoder-side modules are updated. Following
ZipVoice-style flow distillation, we initialize the student from the
pretrained Flow-Matching decoder and distill it into a 4-step generator. Each batch
retains the same two training modes as in the first stage, namely self-reconstruction
and contextual inpainting, and both modes are forwarded during distillation. The learning rate of the first distillation stage is set to 7.5e-5. The learning rate of the second distillation stage is set to 1.5e-5. The number of training steps for each distillation stage is 10k.

For adversarial refinement, we operate directly in mel space with a multi-scale,
multi-band 2D PatchGAN discriminator, rather than a waveform discriminator. The
input is the log-mel spectrogram, and each sample is normalized independently
before being fed into the discriminator. To improve robustness to both local
spectral detail and broader time-frequency structure, we construct four frequency
views: the full band $[0,100]$ and three overlapping sub-bands $[0,32]$,
$[24,72]$, and $[64,100]$. Each view is further evaluated at three temporal
scales $\{1,2,4\}$, yielding 12 mel-domain sub-discriminators in total. We use
least-squares GAN (LSGAN) losses  together with feature matching
losses. Since self-reconstruction samples have different
utterance durations and the supervised regions of contextual inpainting samples
depend on the randomly sampled mask boundary, the effective mel lengths vary
across a mixed batch. We therefore define the adversarial crop length as the
shortest effective mel length $t$ in the batch, and randomly crop one segment of
length $t$ from each real/generated pair before feeding it to the discriminator.

To provide differentiable speaker supervision on generated mel-spectrograms, we train
a proxy speaker encoder to match the embedding space of the frozen WavLM-based speaker
verification model\textsuperscript{\ref{fn:speaker}}. The
resulting mel-level speaker consistency loss is applied only to self-reconstruction
samples. The proxy speaker encoder used in GAN fine-tuning is trained as a mel-to-speaker
embedding regressor, rather than a speaker classifier. Given an input
mel-spectrogram, the model predicts a speaker embedding aligned to the
embedding space of the same frozen WavLM-based speaker verification
model\textsuperscript{\ref{fn:speaker}}. 


The proxy speaker encoder is trained with AdamW using a learning rate of
$1.2\times10^{-4}$. The training objective combines cosine, MSE, and contrastive
distillation terms:
\[
\mathcal{L}_{\text{proxy}} =
\mathcal{L}_{\cos} + 0.1\,\mathcal{L}_{\text{mse}} + 0.1\,\mathcal{L}_{\text{nce}},
\]
where $\mathcal{L}_{\cos}=1-\cos(\hat{s},s)$, $\mathcal{L}_{\text{mse}}$ is the
MSE between normalized predicted and target embeddings, and
$\mathcal{L}_{\text{nce}}$ is an InfoNCE loss.

For each discriminator branch $D_k$, the discriminator loss is:
\[
\mathcal{L}_{D}^{(k)} =
\mathbb{E}\left[(1 - D_k(m_{\text{real}}))^2\right] +
\mathbb{E}\left[D_k(m_{\text{fake}})^2\right].
\]
The generator-side adversarial and feature-matching losses are defined as:
\[
\mathcal{L}_{\text{adv}}^{\text{mel}} =
\sum_k \mathbb{E}\left[(1 - D_k(m_{\text{fake}}))^2\right],
\]
\[
\mathcal{L}_{\text{feat}}^{\text{mel}} =
\sum_k \sum_l
\left\| f_k^{(l)}(m_{\text{fake}}) - f_k^{(l)}(m_{\text{real}}) \right\|_1,
\]
where $f_k^{(l)}(\cdot)$ denotes the $l$-th intermediate feature of the $k$-th
sub-discriminator. The mel-level speaker consistency loss is
\[
\mathcal{L}_{\text{spk}}^{\text{mel}} =
1 - \cos\left(h_{\phi}(\hat{m}), s_{\text{ref}}\right),
\]
where $\hat{m}$ is the mel-spectrogram generated by the 4-step decoder,
$h_{\phi}(\cdot)$ is the proxy speaker encoder, and $s_{\text{ref}}$ is the same
reference speaker embedding used in the first-stage speaker consistency loss. The
final generator objective during GAN fine-tuning is
\[
\mathcal{L}_{G} =
\mathcal{L}_{\text{total}}
+ 0.025\,\mathcal{L}_{\text{adv}}^{\text{mel}}
+ 0.05\,\mathcal{L}_{\text{feat}}^{\text{mel}}
+ 0.5\,\mathcal{L}_{\text{spk}}^{\text{mel}}.
\]
The learning rate of discriminator is set to 1.0e-4. The learning rate of generator is set to 1.5e-5. The total step of gan finetuning stage is 10k.
\subsection{Training Details of Disentanglement Probing}
In the disentanglement probing experiments, tokens extracted from the tokenizers are
first mapped to 128-dimensional embeddings, followed by processing via a two-layer CNN
and a 256-dimensional bidirectional LSTM. For the scenario where multi-layer tokens are
used as input, the token embeddings of all layers are first summed before being
processed by the CNN and LSTM.

For the ASR experiment, English character sequences (including space symbols) are used
as training targets, and the sequential output of the LSTM is optimized against these
sequences using CTC loss. In the speaker classification experiment, the LSTM output is
aggregated by a pooling layer and then mapped to speaker IDs through an MLP layer, with
optimization performed using cross-entropy loss.

For dataset configurations: the LibriSpeech train-960 subset is used for training,
dev-clean for validation, and test-clean for testing in the ASR experiment. In the
speaker classification task, the VoxCeleb1 train, dev, and test subsets serve as the
training, validation, and testing sets, respectively. All models are trained for 30
epochs, with the model achieving the lowest validation WER (for ASR) or the highest
speaker classification accuracy (for speaker classification) selected for testing and
metric calculation. The LibriSpeech training set contains nearly 1000 hours of speech
data, while VoxCeleb1 includes the same 1251 speakers in both its training and testing
sets.

\subsection{Training Details of LLM-based Voice Cloning}
In LLM-based voice cloning tasks, Qwen3-0.6B serves as the LLM backbone. Speech symbols
that depend on the tokenizer’s codebook size and number of layers are first integrated
into the LLM’s vocabulary, with adjustments made to the LLM’s embedding table and
prediction head. We train the LLM using the supervised fine-tuning (SFT) paradigm, where
only the loss of the response part is computed. The model is trained for 5 epochs with a
learning rate of 1e-5 and a warmup ratio of 0.1.

\subsection{Training Details of LLM-based TTS}
In LLM-based voice cloning tasks, Qwen3-0.6B serves as the LLM backbone. Speech symbols
that depend on the tokenizer’s codebook size and number of layers are first integrated
into the LLM’s vocabulary, with adjustments made to the LLM’s embedding table and
prediction head. We train the LLM using the supervised fine-tuning (SFT) paradigm, where
only the loss of the response part is computed. The model is trained for 1 epochs with a learning rate of 5e-5 and a warmup ratio of 0.03.

\section{Baseline Details}
\label{app:baseline_details}
Four representative types of baseline models are selected for fair comparison:
WavTokenizer (single-layer, non-decoupled), SAC and DualCodec (dual-branch, decoupled),
Mimi and Encodec (multi-layer, non-decoupled), and SpeechTokenizer (multi-layer,
decoupled)  Cosy2 $S^3$ (ASR-supervised captures semantics).

\paragraph{WavTokenizer} is built on the VQ-GAN framework and utilizes a single VQ layer
with a codebook size of 4096. Operating at a frame rate of 75 Hz, it achieves a bitrate
of 0.9 kbps.

\paragraph{SAC} is also built on the VQ-GAN framework. It uses a pretrained semantic
tokenizer, while utilizing the DAC framework for the acoustic stream with a total frame
rate of 62.5 Hz. The semantic codebook size is 16,384 and the acoustic codebook size is
16,384, achieving a bitrate of 0.875 kbps.

\paragraph{DualCodec} is a semantic-enhanced tokenizer that directly encodes
semantic-rich SSL features \citep{w2vbert} into its first layer. In our experimental
setting, for a fair comparison, we select the model with six codebook layers that
operates at the frame rate of 12.5Hz. The first codebook contains 16,384 entries, while
the remaining five each contain 4,096 entries, achieving a bitrate of 0.925 kbps.

\paragraph{Mimi} utilizes features from WavLM for semantic distillation. It employs
eight codebooks, each of size 2,048, at a 12.5 Hz frame rate, resulting in a bitrate of
1.1 kbps.

\paragraph{Encodec} is an RVQ-based neural audio codec operating at a frame rate of 75
Hz. In our experimental setting, for a fair comparison, we use only the first two
codebook layers, yielding a bitrate of 1.5 kbps.

\paragraph{SpeechTokenizer} shares a similar architecture with Mini but enhances the
semantic richness of the first-layer tokens via semantic distillation using HuBERT. In
our experimental setting, for a fair comparison, we use only the first two codebook
layers, yielding a bitrate of 1 kbps.

\paragraph{CosyVoice2 $S^3$ Tokenizer} inserts an FSQ module into the encoder of the
SenseVoice-Large ASR model \citep{an2024funaudiollm}, which discretizes intermediate
continuous representations into discrete tokens with textual supervision. This FSQ
module consists of a single-layer FSQ codebook with 8 channels and 3 levels per channel,
resulting in a total of 6561 code entries. With a token rate of 25 Hz, this tokenizer
achieves a bitrate of approximately 0.317 kbps.

\paragraph{Facodec} introduces a novel factorized vector quantization (FVQ) mechanism to
decompose complex speech waveforms into distinct, disentangled subspaces representing
content, prosody, timbre, and acoustic details. It enhances the disentanglement of these
specific attributes through targeted techniques including information bottlenecks,
auxiliary supervision, gradient reversal layers, and detail dropout. In our experimental
setting, we utilize a total of 6 codebook layers—specifically allocating 2 for content,
1 for prosody, and 3 for acoustic details—with a codebook size of 1024, and the timbre
vector.

\section{Additional Experiment Setup Details}
\label{app:exp_setup_details}

\subsection{Dataset Details for Downstream Speech LLMs}
\label{app:llm_data_details}
For LLM-based voice cloning, we construct more than 350,000 training triplets
from the training set of LibriTTS, using f5-tts and cosyvoice2. In each triplet
$(\mathcal{S}, \mathcal{A}, \mathcal{C})$, $\mathcal{S}$ and $\mathcal{A}$ are
taken from different utterances, and $\mathcal{C}$ is synthesized by
two TTS models as the supervised target. For evaluation, we derive more
than 4,000 randomly matched test pairs from LibriTTS test-clean.

For LLM-based TTS, we follow the same data distribution as VoxBox
\citep{sparktts}. This setting is used to study whether different tokenization
schemes provide more effective interfaces for autoregressive speech generation.

\subsection{Downstream Speech LLM Task Formatting}
\label{app:llm_task_format}
For LLM-based voice cloning, each speech utterance is represented by semantic
tokens and acoustic tokens. The autoregressive model is trained on triplets
$(\mathcal{S}, \mathcal{A}, \mathcal{C})$, where $\mathcal{S}$ provides
semantic content, $\mathcal{A}$ provides acoustic style, and
$\mathcal{C}$ denotes the target cloned speech. The model takes the
$[\text{SEP}]$-separated concatenation of $\mathcal{S}$ and $\mathcal{A}$ as
input and predicts the target token sequence $\mathcal{C}$. During inference,
the model generates token sequences for unseen
$(\mathcal{S}', \mathcal{A}')$ pairs, which are then decoded into waveform by
the tokenizer decoder.

For TTS, we adopt a pure LLaSA-style autoregressive architecture
\citep{llasa}. The model generates semantic and acoustic token sequences
conditioned on the input text prompt together with the semantic and acoustic
tokens extracted from the prompt speech. The generated token sequences are then
decoded into speech waveform by the tokenizer decoder.

For both downstream tasks, the semantic and acoustic tokens of DSA-Tokenizer and SAC are 
interleaved in the same sequence.

\subsection{Additional Baseline Notes for Downstream Speech LLMs}
\label{app:llm_baseline_notes}
For both LLM-based voice cloning and TTS, we use Qwen3-0.6B \citep{qwen3} as
the autoregressive backbone and expand its vocabulary with the corresponding
speech tokens.

In LLM-based voice cloning, we compare WavTokenizer, SAC, and DSA-Tokenizer to
analyze how non-disentangled and disentangled token interfaces affect
cross-utterance generation.

In LLM-based TTS, the compared systems include LLaSA-1B-80k, LLaSA-1B-160k,
LLaSA-1B-250k, Spark-TTS, SAC, and DSA-Tokenizer. Among them, the LLaSA and
Spark-TTS systems use non-disentangled tokenizers, while SAC and DSA-Tokenizer
represent disentangled tokenization approaches.

\section{Additional Comparison with the CosyVoice Series}
\label{app:cosyvoice_compare}

We additionally compare DSA-Tokenizer with the CosyVoice series as strong
system-level voice cloning baselines.

\begin{table}[t]
\centering
\scriptsize
\setlength{\tabcolsep}{4pt}
\resizebox{\linewidth}{!}{
\begin{tabular}{lcccccc}
\toprule
\textbf{Model} & UTMOS & WER & SIM & UTMOS$_{zh}$ & CER$_{zh}$ & SIM$_{zh}$ \\
\midrule
CosyVoice1 & 3.55 & 3.24 & 0.40 & 3.00 & 2.94 & 0.60 \\
CosyVoice2 & 3.93 & 3.48 & 0.52 & 3.28 & 2.91 & 0.72 \\
CosyVoice3 & 3.73 & 2.74 & 0.56 & 3.10 & 2.21 & \textbf{0.75} \\
\textbf{DSA-Tokenizer} & \textbf{4.16} & \textbf{2.47} & \textbf{0.61}
& \textbf{3.55} & \textbf{2.16} & 0.71 \\
\bottomrule
\end{tabular}
}
\caption{Additional comparison with the CosyVoice series on
cross-utterance voice cloning.}
\label{tab:cosyvoice_compare}
\end{table}

Table~\ref{tab:cosyvoice_compare} shows that, using continuous
features, the CosyVoice systems are still strong baselines. DSA-Tokenizer
nevertheless outperforms all three CosyVoice systems on English UTMOS, WER,
and SIM. On Chinese, it achieves the best CER and UTMOS while remaining
competitive in SIM.

\section{Disentanglement Probling Result}
The detailed result of disentanglement probing is listed in Table
\ref{tab:disentangle_probing}.

\begin{table}[t]
\centering
\resizebox{\linewidth}{!}{
\begin{tabular}{lcc}
\toprule
\textbf{Model} & \textbf{ASR WER (\%) $\downarrow$} & \textbf{SC ACC (\%) $\uparrow$} \\
\midrule
\textbf{Cosy2 $S^3$ Tokenizer}        &  33.72 & 14.50 \\
\midrule
\textbf{WavTokenizer}                 &  86.31 & 18.14 \\
\midrule
\textbf{EnCodec}                      &        &       \\
\quad Layer 0                         &  64.85 & 16.28 \\
\quad Layer 1-7                       &  69.74 & 17.68 \\
\midrule
\textbf{SpeechTokenizer}              &        &       \\
\quad Layer 0                         &  17.78 & 12.08 \\
\quad Layer 1-7                       &  21.93 & 25.80 \\
\midrule
\textbf{SAC}                          &        &       \\
\quad Semantic token                  &   9.31 &  6.57 \\
\quad Acoustic token                  & 106.24 & 21.82 \\
\midrule
\textbf{Facodec}                &        &       \\
\quad Semantic token                  &   13.5 &  7.13 \\
\quad Acoustic token                  & 100.13 & 22.95 \\
\midrule
\textbf{DSA-Tokenizer}                &        &       \\
\quad Semantic token                  &   6.28 &  2.35 \\
\quad Acoustic token                  & 120.36 & 23.65 \\
\bottomrule
\end{tabular}
}
\caption{Disentanglement probing evaluation results}
\label{tab:disentangle_probing}
\end{table}

\section{Evaluation tools}

The UTMOS score is computed using the pre-trained model available at
\footnote{\url{https://huggingface.co/spaces/sarulab-speech/UTMOS-demo/tree/main}\label{fn:utmos}}.
Speaker similarity (SIM) scores are calculated with a speaker encoder based on WavLM,
fine-tuned for the speaker verification task\ref{fn:speaker}. For English speech, the
word error rate (WER) is evaluated with
Whisper-large-V3\footnote{\url{https://huggingface.co/openai/whisper-large-v3}\label{fn:whisper}},
and for Chinese speech, the character error rate (CER) is evaluated with
Paraformer\footnote{\url{https://huggingface.co/funasr/paraformer-zh}\label{fn:funasr}}.

\section{Broader Impact}

This work introduces a novel dual-stream audio tokenizer capable of disentangling
semantic and acoustic information. Our research holds significant potential for positive
societal impact, particularly in the development of Speech LLMs. The disentangled
representation offers fine-grained control for speech. However, the capability to clone
voices with high similarity raises concerns regarding misuse for deepfakes, unauthorized
voice impersonation, and misinformation.To prevent abuse, we recommend developing robust
deepfake detection tools, speech watermarking, and clear reporting mechanisms.

\section{License Discussion}

The speaker-diarization model\citep{ss} and the model\textsuperscript{\ref{fn:utmos}}
used to calculate the UTMOS score are released under the MIT license.  The speaker
verification fine-tuned WavLM\ref{fn:speaker} is released under the CC BY-SA 3.0
license. Whisper-large-V3\textsuperscript{\ref{fn:whisper}} is released under the Apache
2.0 license. And the model\textsuperscript{\ref{fn:funasr}} calculating CER is released
under model-license license.

The Emilia dataset, LibriSpeech, MagicData-RAMC and LibriTTS are under the license of
cc-by-4.0. GigaSpeech, Libri-Heavy, AISHELL-2, WenetSpeech are released under the Apache
2.0 License. CommonVoice is released under the MPL-2.0 license. MagicData-RAMC and
VoxCeleb1 are under the license CC BY-NC-ND 4.0. People's Speech is under the license of
CC-BY-SA and CC-BY 4.0.

\section{AI Usage}

We used LLMs for grammatical checking and polishing in Sections 1, section 5 and the
Appendix.

\end{document}